**Protecting Privacy and Transforming COVID-19 Case Surveillance Datasets for Public Use**

Authors: Brian Lee, MPH,[1,2] Brandi Dupervil, DHSc, MPH,[1,3] Nicholas P. Deputy, PhD, MPH,[1,3,4] Wil Duck, MPH,[1,5] Stephen Soroka, MPH,[1,6] Lyndsay Bottichio, DrPH, MPH,[1,6] Benjamin Silk, PhD, MPH,[1,4,7] Jason Price,[1,8] Patricia Sweeney, MPH,[1,8] Jennifer Fuld, PhD,[1,9] Todd Weber, MD,[1,6] Dan Pollock, MD[1,6]

Affiliations: [1] COVID-19 response, CDC, Atlanta, GA; [2]Office of the Chief Operations Officer, Office of the Chief Information Officer, CDC, Atlanta, GA; [3]National Center for Birth Defects and Developmental Disabilities, CDC, Atlanta, GA; [4] United States Public Health Service, Rockville, MD; [5]Center for Surveillance, Epidemiology, and Laboratory Services, CDC, Atlanta, GA; [6]National Center for Zoonotic and Emerging Infectious Disease, CDC, Atlanta, GA; [7]National Center for Immunization and Respiratory Diseases, CDC, Atlanta, GA; [8]National Center for HIV/AIDS, Viral Hepatitis, STD, and TB Prevention, CDC, Atlanta, GA; [9]Office of the Associate Director for Policy and Strategy, CDC, Atlanta, GA

Corresponding author: Brian Lee, MPH, Senior Advisor for Informatics, Centers for Disease Control and Prevention, 1600 Clifton Road Northeast, Atlanta, GA, 30329, brian.lee@cdc.gov, 404-498-0046

Disclaimer:  The findings and conclusions in this report are those of the authors and do not necessarily represent the official position of the Centers for Disease Control and Prevention/the Agency for Toxic Substances and Disease Registry.

## Abstract

### Objectives
Federal open data initiatives that promote increased sharing of federally collected data are important for transparency, data quality, trust, and relationships with the public and state, tribal, local, and territorial (STLT) partners. These initiatives advance understanding of health conditions and diseases by providing data to more researchers, scientists, and policymakers for analysis, collaboration, and valuable use outside CDC responders.  This is particularly true for emerging conditions such as COVID-19 where we have much to learn and have evolving data needs. Since the beginning of the outbreak, CDC has collected person-level, de-identified data from jurisdictions and currently has over 8 million records, increasing each day.

This paper describes how CDC designed and produces two de-identified public datasets from these collected data.

### Materials and Methods
Data elements were included based on the usefulness, public request, and privacy implications; specific field values were suppressed to reduce risk of reidentification and exposure of confidential information. Datasets were created and verified for privacy and confidentiality using data management platform analytic tools as well as R scripts.

### Results
Unrestricted data are available to the public through Data.CDC.gov and restricted data, with additional fields, are available with a data use agreement through a private repository on GitHub.com.

## Practice Implications

Enriched understanding of the available public data, the methods used to create these data, and the algorithms used to protect privacy of de-identified individuals allow for improved data use. Automating data generation procedures allows greater and more timely sharing of data.

### 3-question Summary Box (word count: 71/85)
1) What is the current understanding of this subject? Datasets are published by CDC for use by the public, but public users' incomplete understanding of privacy protections and data lineage is a barrier to use.
2) What does this report add to the literature? A description of the fields available for use, locations to access datasets, transformation methods, and privacy protections built into released datasets to reduce risk of re-identification.
3) What are the implications for public health practice? Greater use of released analytical set and improved automation of privacy protections used to create other public health datasets.

## Introduction

Federal open data initiatives that promote increased sharing of federally collected data [1,2,3] are important for transparency, data quality, trust, and relationships with the public and with STLT partners [4]. These initiatives advance understanding of health conditions or diseases by making data available to more researchers, scientists, and policy makers for analyses and other valuable uses. Data sharing initiatives are particularly important during the COVID-19 pandemic, where data needs are constantly evolving and there is much to learn about the disease.

As part of the COVID-19 coordinated response, jurisdictions share de-identified, patient-level data for each case with CDC. These data are sent daily in a combination of three formats – comma separated values (CSV) file, direct data entry of case forms, and National Notifiable Diseases Surveillance System (NNDSS) electronic case notifications – to CDC's Data Collation and Integration for Public Health Event Response (DCIPHER) system. DCIPHER is a data management and analysis system using Palantir Foundry [5] software that allows analysis via R [6], Python [7], and an analytic tool called Contour. The data are managed with a DCIPHER Case Surveillance Pipeline, a series of linked programs that cleans, collates, de-duplicates, and transforms data to produce an analytical-ready epidemiological dataset used across the response. Data do not include direct identifiers but do include demographic characteristics, exposure history, disease severity indicators and outcomes, clinical data, laboratory diagnostic test results, and comorbidities (Figure 1).

CDC's Case Surveillance Section, the response group established to conduct surveillance activities and serve as the data steward over case data, created a new process that transforms the epidemiological dataset with privacy protection algorithms to systematically create anonymized subset data. This process contains automated workflows and R statistical software (version 4.0.3; The R Foundation) that implement and validate field level suppression for k-anonymity [8] and l-diversity [9] levels to release microdata, monthly, in two public datasets:

- COVID-19 Case Surveillance Public Use Data - a "public use" dataset designed with 11 data fields is accessible via Data.CDC.gov [10], with an interactive visualization to allow the public to filter, sort, and perform exploratory analysis.
- COVID-19 Case Surveillance Restricted Access Detailed Data - a "scientific use" dataset designed with 31 fields, and more stringent privacy protections, to provide more detailed information for scientists, statisticians, journalists, and researchers; however, this requires users to sign a Registration Information and Data Use Restriction Agreement (RIDURA) to access through a private GitHub repository [11].

To increase usability and foster transparency, this paper describes dataset definitions, design of the pipeline that creates them, and privacy protection rationale.

## Materials and Methods

Multiple groups within CDC's emergency response organization worked together to design the public case datasets. Surveillance Review and Response Group (SRRG), a group established to improve data use within the response, worked with the Case Surveillance Section to use privacy heuristics, available guidance, and codes of practice [12,13] to design the Data Sharing Privacy Review Procedures, a seven-step process (Figure 2) that implemented CDC's data release policies [14] to protect privacy and publish useful and accessible data. This privacy review process was used to derive two datasets from the epidemiological dataset (Figure 1). Data elements were selected for inclusion in both datasets based on usefulness, public request, and privacy implications. Specific field values (e.g., *age_group*, *race_ethnicity_combined*) were suppressed to reduce risk of re-identification and exposure of confidential information. Datasets were created and verified for privacy and confidentiality standards using Palantir Contour and R scripts [15] using the sdcMicro package [16].

The privacy procedures reduce the risk of re-identifying patients while providing useful information. To meet these privacy protection needs, not all variables can be released. Since the public use dataset is widely accessible, its data are the most restricted while the scientific use dataset is released only to approved researchers and includes more variables.

Re-identification risk cannot be reduced to zero, but this systematic process is designed to make this risk low [17] to protect individuals whose data contributes to these public datasets.

Step 1: Classify Variables. All variables from the epidemiological dataset were reviewed and classified according to their sensitivity into one of four categories: direct identifiers, quasi-identifiers, confidential attributes, and non-confidential attributes. Direct identifiers are variables that would unambiguously identify an individual (e.g., name, address) and while CDC does not receive these types of data, each field is checked and confirmed that no identifying information is contained in an open-ended or free text response. Quasi-identifiers are fields that may identify an individual if they occur rarely enough in a dataset or could be combined with other fields or data (e.g., age group, sex, county). Confidential attributes are sensitive information that would not commonly be known about an individual (e.g., first positive specimen date). Non-confidential attributes are general information that cannot be used to identify individuals but still may potentially be combined with other data (e.g., case status). Fields are reviewed individually and as a combined set of fields within the dataset. From this review, all potential fields were either included, excluded, or transformed to reduce sensitivity. For example, *date_of_birth* is excluded, so we created a generalized *age_group* field using ten-year bins with a top-coded bin for 80+.

We finalized the design of the datasets by identifying the specific fields included in the public use (Supplement 1) and scientific use (Supplement 2) datasets. Fields were identified by evaluating their analytical usefulness with their re-identification risk. Fields included in the datasets were adjusted over time incorporating feedback. For example, additional geographic fields for county were added to the scientific use dataset; and race/ethnicity was added to the public use dataset. Geographic fields were only included in the scientific use dataset by researchers who sign a data use agreement.

For the public use dataset, the most current dataset contains 11 fields with three quasi-identifier fields – *sex*, *age_group*, *race_ethnicity_combined* – and one confidential attribute – *pos_spec_date*. For the scientific use dataset, the dataset includes 31 fields, with six quasi-identifiers – *sex*, *age_group*, *race_ethnicity_combined*, *res_county*, *res_state*, *hc_work_yn* – and one confidential attribute – *pos_spec_date*. These fields are used in subsequent steps to establish and check cell suppression levels (Table 1).

Step 2: Review for Personally Identifiable Information (PII). We reviewed all data fields to confirm that no PII was present. All data fields were limited to categorical, date, and numeric values and were reviewed and confirmed that they could not contain PII. All free text data fields that had the potential to contain PII were excluded.

Step 3: Set Privacy Levels. We established privacy thresholds by defining the minimum acceptable size for the number of records in the dataset that share quasi-identifiers. Although there is no universal threshold [18], a minimum level is suggested, with a common recommendation of 5 and uncommonly above 5 [19]. We set this level at 5 to be conservative and consistent with previous approaches used in public health [20]. This means that no fewer than 5 records are allowed to share values from a single, or any combination of, quasi-identifier fields. Workflows called "Contour boards" were created in DCIPHER to automatically detect any combination of quasi-identifiers meeting our criteria for small cell suppression and set fields to "NA". Only field values were suppressed, records remained in the dataset so researchers can identify when suppression criteria were applied. When suppressing fields, data managers made every effort to suppress as few fields as possible while meeting the privacy level (Table 1).

Step 4: Re-code Variables. We used common variable coding techniques within the pipeline to clean and ensure uniformity of the responses within each field. Questions that were left unanswered in the case report form were re-classified to "Missing", with the following exceptions: *age_group* recoded to "Unknown;" *res_state* recoded to the reporting jurisdiction; and *res_county* and *county_fips_code* were left unchanged. Logic checks were performed on dates to detect illogical responses and set them to "Null" until the jurisdiction updates; for example, dates reported in the future, or dates reported prior to the onset of COVID. Additionally, initial COVID report date was examined and when the value was blank upon receipt from the reporting jurisdiction, the value was set to the date the data file was first submitted to CDC. The primary goal was to ensure consistency in applying suppression and to simplify the dataset for ease of use and analysis.

Case-based surveillance data are dynamic and jurisdictions can modify and re-submit when new information becomes available; therefore, records may change between releases with de-deduplicated updates. Data are only included in public datasets with a 14-day delay based on the *cdc_report_dt* field. This allows data managers time to review responses and work with jurisdictions to correct data quality issues. The original release on May 18[th] used a 30-day window, but was updated in subsequent updates as improved data quality reviews showed minimal changes after 14 days.

Step 5: Review k-Anonymity. Each time datasets are generated, we review for k-anonymity. K-anonymity is a technique used to reduce the risk of re-identifying a person or linking person-specific data to other information based on a rare combination of quasi-identifiers. We use this technique to suppress quasi-identifiers values so that each person contained in the released dataset cannot be distinguished from at least *k-1* other persons who share the same quasi-identifiers [8]. This technique uses privacy thresholds established in step 3 across all quasi-identifiers classified in step 1.

Figure 2 shows an example of how k-anonymity is used to suppress record quasi-identifier values using only 10 records to illustrate how k-anonymity applies to the entirety of both datasets. Fields on the left are the raw data before suppression. The *frequency* field indicates the number of records in the example dataset that have the same combination of quasi-identifiers; for example, the first record *frequency* =1, meaning that the combination of *sex*, *age_group*, and *race_ethnicity_combined* quasi-identifiers occurs only once within the data. Since we require 5-anonymity, we will suppress fields so that their quasi-identifiers occur at least 5 times. After suppression *frequency* shows that each record's quasi-identifiers occurs 5 times. Note that records are never removed, and, in this example, since we suppress the fewest fields possible to create a cell with 5 members, only *sex* and *race_ethnicity_combined* were suppressed and we were able to leave *age_group* unchanged. This example includes the three quasi-identifiers within the public use dataset but functions the same for the scientific use dataset using its six quasi-identifiers.

After every time datasets are regenerated through the pipeline, data managers use R programs [15] to verify that each generated dataset meets the levels established in step 3. If any errors are detected, the pipeline is revised to correct the bug and the datafile is regenerated and retested until both processes are satisfied. At the end of this step, each dataset is verified to be 5-anonymous.

The number of times fields are suppressed within each dataset varies with each monthly release and between datasets because suppression depends on the total number of rows in the dataset and on the number of included fields (Supplement 3, Supplement 4). Users should consider the amount of suppression within fields as they design and create analyses.

Step 6: Review l-Diversity. As an extension of the k-anonymity check, step 6 involves checking for l-diversity to reduce the risk of exposing confidential information on an individual. L-diversity, another technique to protect confidential information, checks to ensure that for a group of individuals who share the same quasi-identifiers, at least *l* distinct values exist for each confidential attribute [9]. These datasets require 2-diversity so that confidential variables cannot be determined in situations where records share the same quasi-identifiers values.

Figure 3 shows an example of how l-diversity is used to suppress specific confidential values within records to meet the privacy levels. Again, the fields on the left are raw data. The *distinct* field indicates the number of unique *pos_spec_dt* confidential field values shared by all records with the same quasi-identifiers; notice that some records have a *distinct* of 1 because they all share the same *sex* field value of "Female," *age_group* field value of "0-9," and *race_ethnicity_combined* field value "Asian, Non-Hispanic" and all share the same *pos_spec_dt* value of "2020-03-31." Since our requirement for the dataset is that we must have 2-diversity, the confidential field is suppressed and set to "NA" to not reveal the *pos_spec_dt* value. The distinct value remains 1 but now the value is "NA" and cannot be known. This prevents someone knowing the specific specimen date just because they know the person's sex, age group, race, and ethnicity. Records are never removed, only field values are suppressed.

Step 7: Research Links. Finally, in step 7, to reduce the risk of "mosaic effect" [1], we researched other publicly available datasets that could be linked by quasi-identifiers to identify individuals. The mosaic effect is a risk where information within an individual dataset may not identify an individual, but when combined with other available datasets may have a risk to identify individuals. This risk is reduced through the use of k-anonymity levels that reduce the number of rare combinations of quasi-identifiers that could be linked to other datasets; however, it is challenging to completely eliminate this risk. We reviewed the 13 COVID-19 related datasets published on Data.CDC.gov at that time [20]. We were not able to exhaustively search all available datasets but did review quasi-identifiers against the other 543 datasets, at that time, published by CDC with machine-readable metadata available through the Data.CDC.gov public data catalog [22].

## Results

The public use dataset, updated monthly, was published to https://data.cdc.gov/Case-Surveillance/COVID-19-Case-Surveillance-Public-Use-Data/vbim-akqf on May 18, originally contained 339,301 records with 9 fields. On May 29 we added *onset_dt*. On June 27 we added

*race_ethnicity_combined*. As of December 4, it contains 8,405,079 records, every case through November 19 (Supplement 3). To support the most users, CDC releases these data following the FAIR Guiding Principles of findability, accessibility, interoperability and reusability [24], using machine-readable CSV formats and an open standards-compliant application programming interface. The dataset was viewed over 438,000 times and downloaded over 24,000 times (Supplement 5). An interactive visualization, https://data.cdc.gov/Case-Surveillance/COVID-19-Case-Surveillance-Public-Use-Data-Profile/xigx-wn5e, was used over 15,000 times. Google Scholar shows twenty-five publications referencing these data [25]. Source code is available at https://github.com/cdcgov/covid_case_privacy_review.

The scientific use dataset, updated monthly, was published to a private GitHub repository on May 18, containing 315,593 records with 29 fields. On June 27 we updated the dataset to 31 fields: combined *race* and *ethnicity* into *race_ethnicity_combined* and added *res_state*, *res_county*, and *county_fips_code*. As of December 4, it contained 8,405,079 records representing every case received by CDC through November 19 (Supplement 4). GitHub is a third-party web site that CDC uses to make data easier for researchers to download datafiles as industry-standard, zip-compressed, CSV files. Dataset descriptions and RIDURA instructions are available at https://data.cdc.gov/Case-Surveillance/COVID-19-Case-Surveillance-Restricted-Access-Detai/mbd7-r32t. The dataset had been accessed by 94 researchers as of December 11, 2020, and Google Scholar shows two papers referencing these data [23].

## Discussion

Public datasets are needed for open government and transparency; promotion of research; and efficiency. Specifically, COVID-19 case data transparency is important for fostering and maintaining trust and relationships with the public and STLT public health partners [26]. To balance the need to create and share public use datasets with protection of patients' privacy and confidential information, we created a seven-step data sharing privacy review to protect privacy and publish useful data.

There are a large number and variety of repositories for public datasets [27]. Given the large number of repositories for public datasets, and the large number of datasets contained within each repository, we were unable to develop a practical, systematic process to review all public datasets and ensure with complete certainty that the risk of re-identifying patients in our datasets through the use of quasi-identifiers is completely eliminated. For example, a single popular repository for public research data, figshare.com, revealed 803 results for "COVID", illustrating the large number of datasets that exist.

We compensated for this by reducing the number of variables, generalizing variables, and establishing conservative k-anonymity levels. As methods improve to compare data to other released datasets to rule out security concerns, we could include additional fields or apply more precise privacy levels making the data more useful for analysis.

## Practice Implications

Systematic privacy review procedures are important for data engineering purposes to collaborate on and validate data design across systems, locations, and teams. Privacy review is complex, and requirements must be understood by epidemiologists, statisticians, data product owners, informaticians, analysts, health communicators, and data custodians so that they are implemented, tested, and applied reliably each time that a dataset is updated. Automated computational privacy controls are important to meet the volume and schedule of data updates while reliably meeting privacy requirements as this is not possible with manual processes.

Release of these datasets has led to improved data quality by incorporating user feedback into continual improvements of the data pipeline for public and non-public data, such as consistently coding missing values, adding county coding, and more accurately identifying state and county of residence. Public data are part of the data feedback loop throughout the data lifecycle where more users of data are able to identify and prioritize data features and bugfixes.

Through the creation of these datasets and implementation of computational privacy protections, CDC contributed to a knowledge base of COVID-19 data practices that will be used for design and publication of additional datasets beyond case surveillance. CDC publishes 40 different COVID-19 public datasets on Data.CDC.gov [28] as of November 18, 2020. Currently two datasets use these computational privacy protections; additional datasets will be published based on feedback and public health program priority.

These case datasets are now available to the public for review, use in research, and to improve data transparency with partners. The practices and tools developed to design and release these data are available to other programs within CDC's COVID-19 response through the shared data pipeline, privacy review procedures maintained by SRRG, and computational privacy review software. With increased, systematic releases of these public datasets and more training and information available, we expect increased use and greater public health benefit.

## Tables and Figures

Figure 1. Case Surveillance Data Flow Process Includes Specific Pipelines to Implement Unique Privacy Protections for each Public Dataset

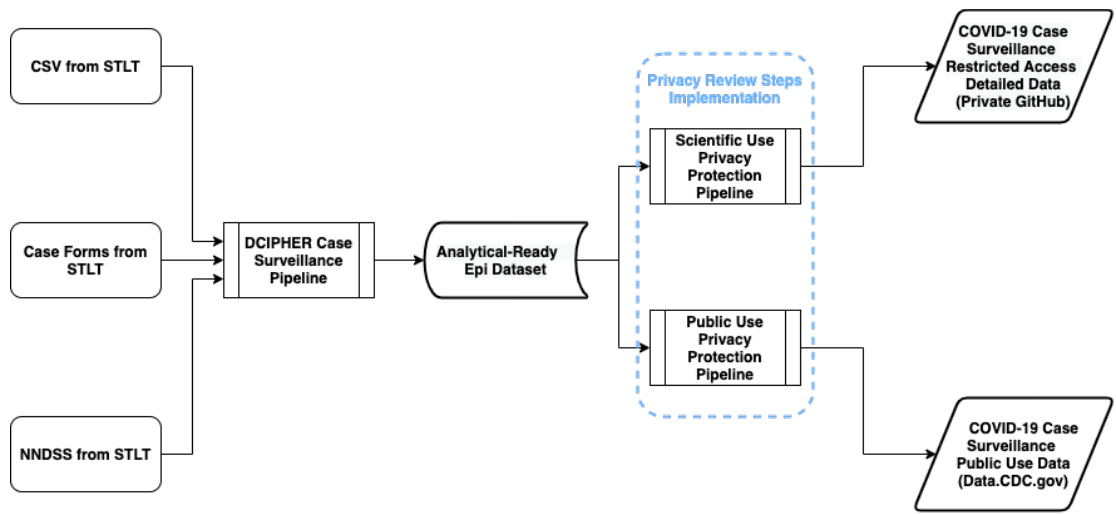

CSV = Comma-separated Values file; STLT = State, Tribal, Local, Territorial public health jurisdictions; NNDSS = National Notifiable Diseases Surveillance System; DCIPHER = Data Collation and Integration for Public Health Event Response system

Figure 2. Privacy Review Steps Used in Designing Public Datasets (PII = Personally Identifiable Information)

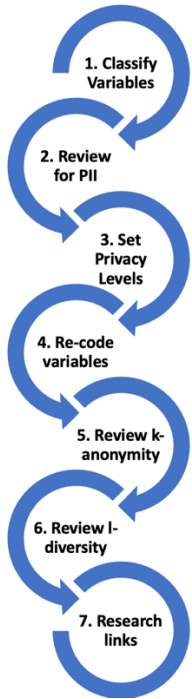

Figure 3. Example of k-anonymity Field Level Suppression for Quasi-Identifiers

| Raw | | | | | Suppressed | | | |
|---|---|---|---|---|---|---|---|---|
| sex | age_group | race_ethnicity_combined | frequency | | sex | age_group | race_ethnicity_combined | frequency |
| Male | 0-9 | Hispanic/Latino | 1 | | NA | 0-9 | NA | 5 |
| Female | 0-9 | Hispanic/Latino | 1 | | NA | 0-9 | NA | 5 |
| Unknown | 0-9 | Hispanic/Latino | 5 | | Unknown | 0-9 | Hispanic/Latino | 5 |
| Male | 0-9 | Unknown | 1 | | NA | 0-9 | NA | 5 |
| Female | 0-9 | Unknown | 1 | => | NA | 0-9 | NA | 5 |
| Unknown | 0-9 | Unknown | 1 | | NA | 0-9 | NA | 5 |
| Unknown | 0-9 | Hispanic/Latino | 5 | | Unknown | 0-9 | Hispanic/Latino | 5 |
| Unknown | 0-9 | Hispanic/Latino | 5 | | Unknown | 0-9 | Hispanic/Latino | 5 |
| Unknown | 0-9 | Hispanic/Latino | 5 | | Unknown | 0-9 | Hispanic/Latino | 5 |
| Unknown | 0-9 | Hispanic/Latino | 5 | | Unknown | 0-9 | Hispanic/Latino | 5 |

Figure 4. Example of l-diversity Field Suppression for Confidential Attributes

| Raw | | | | | | Suppressed | | | | |
|---|---|---|---|---|---|---|---|---|---|---|
| sex | age_group | race_ethnicity_combined | pos_spec_dt | distinct | | sex | age_group | race_ethnicity_combined | pos_spec_dt | distinct |
| Female | 0-9 | Asian, Non-Hispanic | 2020-03-01 | 1 | | Female | 0-9 | Asian, Non-Hispanic | NA | 1 |
| Female | 0-9 | Asian, Non-Hispanic | 2020-03-01 | 1 | | Female | 0-9 | Asian, Non-Hispanic | NA | 1 |
| Unknown | 0-9 | Hispanic/Latino | 2020-04-01 | 4 | | Unknown | 0-9 | Hispanic/Latino | 2020-04-01 | 4 |
| Female | 0-9 | Asian, Non-Hispanic | 2020-03-01 | 1 | | Female | 0-9 | Asian, Non-Hispanic | NA | 1 |
| Female | 0-9 | Asian, Non-Hispanic | 2020-03-01 | 1 | => | Female | 0-9 | Asian, Non-Hispanic | NA | 1 |
| Female | 0-9 | Asian, Non-Hispanic | 2020-03-01 | 1 | | Female | 0-9 | Asian, Non-Hispanic | NA | 1 |
| Unknown | 0-9 | Hispanic/Latino | 2020-05-01 | 4 | | Unknown | 0-9 | Hispanic/Latino | 2020-05-01 | 4 |
| Unknown | 0-9 | Hispanic/Latino | 2020-05-01 | 4 | | Unknown | 0-9 | Hispanic/Latino | 2020-05-01 | 4 |
| Unknown | 0-9 | Hispanic/Latino | 2020-06-01 | 4 | | Unknown | 0-9 | Hispanic/Latino | 2020-06-01 | 4 |
| Unknown | 0-9 | Hispanic/Latino | 2020-07-01 | 4 | | Unknown | 0-9 | Hispanic/Latino | 2020-07-01 | 4 |

Table 1. Variables Classified as Quasi-Identifier or Confidential Fields, by Type of Public Dataset (k is threshold for k-anonymity, l is the threshold for l-diversity)

| Public Use Dataset (11 fields) | Scientific Use Dataset (31 fields) |
|---|---|
| Privacy thresholds: k=5, l=2 | Privacy thresholds: k=5, l=2 |
| Quasi-identifier fields (3):<br>• sex<br>• age_group<br>• race_ethnicity_combined | Quasi-identifier fields (6):<br>• sex<br>• age_group<br>• race_ethnicity_combined<br>• res_county<br>• res_state<br>• hc_work_yn |
| Confidential fields (1):<br>• pos_spec_dt | Confidential fields (1):<br>• pos_spec_dt |

Supplement 1. CDC COVID-19 Case Surveillance Public Use Data Dictionary (11 variables)

| Description | Variable | Source | Values | Type | Calculation (if applicable) |
|---|---|---|---|---|---|
| Date case was first reported to the CDC | cdc_report_dt | Calculated | YYYY-MM-DD | Date | This date was populated using the date at which a case record was first submitted to the database. If missing, then the report date entered on the case report form was used. If missing, then the date at which the case first appeared in the database was used. |
| Date of first positive specimen collection | pos_spec_dt | Case Report Form | YYYY-MM-DD | Date | NA, if value suppressed for privacy protection. |
| Date of symptom onset | onset_dt | Case Report Form | YYYY-MM-DD | Date | |
| What is the current status of this person? | current_status | Case Report Form | Laboratory-confirmed case<br>Probable Case | Character | |
| Sex | sex | Case Report Form | Male<br>Female<br>Unknown<br>Missing<br>Other | Character | NA, if value suppressed for privacy protection. |

| Age group | age_group | Calculated | 0 - 9 Years<br>10 - 19 Years<br>20 - 39 Years<br>40 - 49 Years<br>50 - 59 Years<br>60 - 69 Years<br>70 - 79 Years<br>80 + Years<br>Unknown | Character | The age group categorizations were populated using the age value that was reported on the case report form. Date of birth was used to fill in missing/unknown age values using the difference in time between date of birth and onset date.<br>NA, if value suppressed for privacy protection. |
|---|---|---|---|---|---|
| Race and Ethnicity (combined) | race_ethnicity_combined | Calculated | American Indian/Alaska Native, Non-Hispanic<br>Asian, Non-Hispanic<br>Black, Non-Hispanic<br>Multiple/Other, Non-Hispanic<br>Native Hawaiian/Other Pacific Islander, Non-Hispanic<br>White, Non-Hispanic<br>Hispanic/Latino<br>Unknown<br>Missing | Character | If more than race was reported, race was categorized into multiple/other races.<br>NA, if value suppressed for privacy protection. |
| Was the patient hospitalized? | hosp_yn | Case Report Form | Yes<br>No<br>Unknown<br>Missing | Character | |
| Was the patient admitted to an intensive care unit (ICU)? | icu_yn | Case Report Form | Yes<br>No<br>Unknown<br>Missing | Character | |
| Did the patient die as a result of this illness? | death_yn | Case Report Form | Yes<br>No<br>Unknown<br>Missing | Character | |

| Pre-existing medical conditions? | medcond_yn | Case Report Form | Yes<br>No<br>Unknown<br>Missing | Character | |

Supplement 2. CDC COVID-19 Case Surveillance Restricted Access Detailed Data Dictionary (31 variables)

| Description | Variable | Source | Values | Type | Calculation (if applicable) |
|---|---|---|---|---|---|
| What is the current status of this person? | current_status | Case Report Form | Laboratory-confirmed case<br>Probable Case | Character | |
| Date case was first reported to the CDC | cdc_report_dt | Calculated | YYYY-MM-DD | Date | This date was populated using the date at which a case record was first submitted to the database. If missing, then the report date entered on the case report form was used. If missing, then the date at which the case first appeared in the database was used. |
| Sex | sex | Case Report Form | Male<br>Female<br>Unknown<br>Other<br>Missing | Character | NA, if value suppressed for privacy protection. |
| Age group | age_group | Calculated | 0 - 9 Years<br>10 - 19 Years<br>20 - 39 Years<br>40 - 49 Years<br>50 - 59 Years<br>60 - 69 Years<br>70 - 79 Years<br>80 + Years<br>Unknown | Character | The age group categorizations were populated using the age value that was reported on the case report form. Date of birth was used to fill in missing/unknown age values using the difference in time between |

| | | | | | date of birth and onset date. NA, if value suppressed for privacy protection. |
|---|---|---|---|---|---|
| Race and Ethnicity (combined) | race_ethnicity_combined | Calculated | American Indian/Alaska Native, Non-Hispanic Asian, Non-Hispanic Black, Non-Hispanic Multiple/Other, Non-Hispanic Native Hawaiian/Other Pacific Islander, Non-Hispanic White, Non-Hispanic Hispanic/Latino Unknown Missing | Character | If more than race was reported, race was categorized into multiple/other races. NA, if value suppressed for privacy protection. |
| County Federal Information Processing Standards (FIPS) Code | county_fips_code | Calculated | | Character | FIPS Code was derived based on resident state and resident county. NA, if value suppressed for privacy protection or missing. |
| County of residence | res_county | Case Report Form | | Character | NA, if value suppressed for privacy protection or missing. |
| State of residence | res_state | Case Report Form | | Character | NA, if value suppressed for privacy protection. |
| Date of symptom onset | onset_dt | Case Report Form | YYYY-MM-DD | Date | |
| Date of first positive | pos_spec_dt | Case Report Form | YYYY-MM-DD | Date | NA, if value suppressed for |
| specimen collection | | | | | privacy protection. |
| Was the patient hospitalized? | hosp_yn | Case Report Form | Yes<br>No<br>Unknown<br>Missing | Character | |
| Was the patient admitted to an intensive care unit (ICU)? | icu_yn | Case Report Form | Yes<br>No<br>Unknown<br>Missing | Character | |
| Did the patient die as a result of this illness? | death_yn | Case Report Form | Yes<br>No<br>Unknown<br>Missing | Character | |
| Is the patient a health care worker in the United States? | hc_work_yn | Case Report Form | Yes<br>No<br>Unknown<br>Missing | Character | NA, if value suppressed for privacy protection. |
| Did the patient develop pneumonia? | pna_yn | Case Report Form | Yes<br>No<br>Unknown<br>Missing | Character | |
| Did the patient have an abnormal chest X-ray? | abxchest_yn | Case Report Form | Yes<br>No<br>Unknown<br>Missing<br>N/A | Character | |
| Did the patient have acute respiratory distress syndrome? | acuterespdistress_yn | Case Report Form | Yes<br>No<br>Unknown<br>Missing | Character | |
| Did the patient receive mechanical ventilation (MV)/intubation? | mechvent_yn | Case Report Form | Yes<br>No<br>Unknown<br>Missing | Character | |
| Fever >100.4F (38C) | fever_yn | Case Report Form | Yes<br>No<br>Unknown<br>Missing | Character | |
| Subjective fever (felt feverish) | sfever_yn | Case Report Form | Yes<br>No<br>Unknown<br>Missing | Character | |
| Chills | chills_yn | Case Report Form | Yes<br>No<br>Unknown<br>Missing | Character | |
| Muscle aches (myalgia) | myalgia_yn | Case Report Form | Yes<br>No<br>Unknown<br>Missing | Character | |
| Runny nose (rhinorrhea) | runnose_yn | Case Report Form | Yes<br>No<br>Unknown<br>Missing | Character | |
| Sore throat | sthroat_yn | Case Report Form | Yes<br>No<br>Unknown<br>Missing | Character | |
| Cough (new onset or worsening of chronic cough) | cough_yn | Case Report Form | Yes<br>No<br>Unknown<br>Missing | Character | |
| Shortness of breath (dyspnea) | sob_yn | Case Report Form | Yes<br>No<br>Unknown<br>Missing | Character | |
| Nausea or Vomiting | nauseavomit_yn | Case Report Form | Yes<br>No<br>Unknown<br>Missing | Character | |
| Headache | headache_yn | Case Report Form | Yes<br>No<br>Unknown<br>Missing | Character | |
| Abdominal pain | abdom_yn | Case Report Form | Yes<br>No<br>Unknown<br>Missing | Character | |
| Diarrhea (≥3 loose/looser than normal stools/24hr period) | diarrhea_yn | Case Report Form | Yes<br>No<br>Unknown<br>Missing | Character | |
| Pre-existing medical conditions? | medcond_yn | Case Report Form | Yes<br>No<br>Unknown<br>Missing | Character | |

Supplement 3. Suppression Summary for Public Use Dataset, December 4 Release (n=8,405,079)

| Field Name | Number of Values per Field Suppressed | Percent of Values per Field Suppressed |
|---|---|---|
| race_ethnicity_combined | 7 | 0.00% |
| sex | 18 | 0.00% |
| age_group | 89 | 0.00% |

Supplement 4. Suppression Summary for Scientific Use Dataset, December 4 Release (n=8,405,079)

| Field Name | Number of Values per Field Suppressed | Percent of Values per Field Suppressed |
| --- | --- | --- |
| race_ethnicity_combined | 272,207 | 3.24% |
| sex | 94,591 | 1.13% |
| age_group | 46,485 | 0.55% |
| hc_work_yn | 317,355 | 3.78% |
| res_county | 284,499 | 3.38% |
| res_state | 1,921 | 0.02% |

Supplement 5. Access and Use by the Public of the CDC COVID-19 Case Surveillance Public Use Data Increased Through the Response [Combined Access types from May 19, 2020 through November 24, 2020]

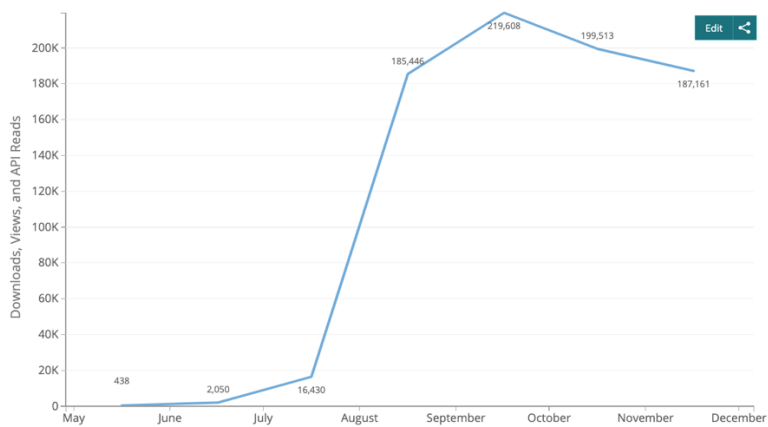

## Acknowledgements

Marion Anandappa, MS; Amy Baugher, MPH; Katie Fullerton, MPH; Jeremy Gold, MD, MS; Chad Heilig, PhD; Sara Johnston, MPH; Beth Pallo, MPH; Paula Yoon, ScD, MPH; CDC COVID-19 Response Case Surveillance Section; CDC COVID-19 Response Surveillance Review and Response Group.

## Ethics declarations

The authors declare no competing interests.

## References

[1] Office of Management and Budget. (2013). Open Data Policy-Managing Information as an Asset (OMB Memorandum M-13-13). https://digital.gov/open-data-policy-m-13-13/

[2] Office of Management and Budget. (2009, December 8). Open Government Directive (OMB Memorandum M-10-06). The White House. https://obamawhitehouse.archives.gov/node/7011

[3] Secretary (IOS), I. O. of the. (2015, January 26). Open Government at HHS [Text]. HHS.Gov. https://www.hhs.gov/open/index.html

[4] AbouZahr, C., Adjei, S., & Kanchanachitra, C. (2007). From data to policy: good practices and cautionary tales. Lancet (London, England), 369(9566), 1039–1046. https://doi.org/10.1016/S0140-6736(07)60463-2

[5] Palantir Foundry. (n.d.). Palantir. Retrieved November 24, 2020, from https://palantir.com/palantir-foundry/

[6] R Core Team. (n.d.). R: A Language and Environment for Statistical Computing (3.5) [Computer software]. R Foundation for Statistical Computing. Retrieved November 24, 2020, from https://www.R-project.org/

[7] Python Software Foundation. (n.d.). Python Language Reference (3.6) [Computer software]. Retrieved November 24, 2020, from https://www.python.org/

[8] Samarati, P., & Sweeney, L. (1998). Protecting privacy when disclosing information: k-anonymity and its enforcement through generalization and suppression.

[9] Machanavajjhala, A., Kifer, D., Gehrke, J., & Venkitasubramaniam, M. (2007). L-diversity: Privacy beyond k-anonymity. ACM Transactions on Knowledge Discovery from Data, 1(1), 3. https://doi.org/10.1145/1217299.1217302

[10] COVID-19 Case Surveillance Public Use Data | Data | Centers for Disease Control and Prevention. (n.d.). Retrieved November 18, 2020, from https://data.cdc.gov/Case-Surveillance/COVID-19-Case-Surveillance-Public-Use-Data/vbim-akqf

[11] COVID-19 Case Surveillance Restricted Access Detailed Data | Data | Centers for Disease Control and Prevention. (n.d.). Retrieved November 25, 2020, from https://data.cdc.gov/Case-Surveillance/COVID-19-Case-Surveillance-Restricted-Access-Detai/mbd7-r32t

[12] US Department of Health and Human Services. (2012, September 7). Guidance Regarding Methods for De-identification of Protected Health Information in Accordance with the Health Insurance Portability and Accountability Act (HIPAA) Privacy Rule [Text]. HHS.Gov. https://www.hhs.gov/hipaa/for-professionals/privacy/special-topics/de-identification/index.html

[13] Federal Committee on Statistical Methodology. (2005). Statistical Policy Working Paper 22, Report on Statistical Disclosure Limitation Methodology. US Federal Committee on Statistical Methodology. https://nces.ed.gov/fcsm/pdf/spwp22.pdf

[14] Centers for Disease Control and Prevention, Agency for Toxic Substances and Disease Registry. (2016). Policy on Public Health Research and Nonresearch Data Management and Access. Centers for Disease Control and Prevention, Department of Health and Human Services. https://www.cdc.gov/maso/policy/policy385.pdf [15] COVID-19 Case Privacy Review. (n.d.). Centers for Disease Control and Prevention. Retrieved November 18, 2020, from https://github.com/cdcgov/covid_case_privacy_review [16] Templ, M., Kowarik, A., & Meindl, B. (2015).

Statistical Disclosure Control for Micro-Data Using the R Package sdcMicro. Journal of Statistical Software, 67(4). https://doi.org/10.18637/jss.v067.i04 [17] El Emam, K., Jonker, E., Arbuckle, L., & Malin, B. (2011). A Systematic Review of Re-Identification Attacks on Health Data. PLoS ONE, 6(12). https://doi.org/10.1371/journal.pone.0028071 [18] Committee on Strategies for Responsible Sharing of Clinical Trial Data; Board on Health Sciences Policy; Institute of Medicine. (2015). Concepts and Methods for De-identifying Clinical Trial Data. National Academies Press (US). https://www.ncbi.nlm.nih.gov/books/NBK285994/ [19] El Emam, K., & Dankar, F. K. (2008). Protecting Privacy Using k-Anonymity. Journal of the American Medical Informatics Association : JAMIA, 15(5), 627–637. https://doi.org/10.1197/jamia.M2716 [20] Integrated Guidance for Developing Epidemiologic Profiles: HIV Prevention and Ryan White HIV/AIDS Programs Planning. (2014). Centers for Disease Control and Prevention and Health Resources and Services Administration. https://www.cdc.gov/hiv/pdf/guidelines_developing_epidemiologic_profiles.pdf

[21] Search & Browse covid-19 | Data | Centers for Disease Control and Prevention. (n.d.). Retrieved May 10, 2020, from https://data.cdc.gov/browse?tags=covid-19

[22] Search & Browse Page 1 of 94 | Data | Centers for Disease Control and Prevention. (n.d.). Retrieved November 18, 2020, from https://data.cdc.gov/browse

[23] Search for suggested citation "COVID-19 Case Surveillance Data Access, Summary, and Limitations." (n.d.). Google Scholar. Retrieved November 18, 2020, from https://scholar.google.com/scholar?hl=en&as_sdt=0%2C11&as_ylo=2020&q=%22COVID-19+Case+Surveillance+Data+Access%2C+Summary%2C+and+Limitations%22&btnG=

[24] Wilkinson, M. D., Dumontier, M., Aalbersberg, Ij. J., Appleton, G., Axton, M., Baak, A., Blomberg, N., Boiten, J.-W., da Silva Santos, L. B., Bourne, P. E., Bouwman, J., Brookes, A. J., Clark, T., Crosas, M., Dillo, I., Dumon, O., Edmunds, S., Evelo, C. T., Finkers, R., … Mons, B. (2016). The FAIR Guiding Principles for scientific data management and stewardship. Scientific Data, 3(1), 1–9. https://doi.org/10.1038/sdata.2016.18

[25] Search for dataset url fragment "data.cdc.gov/Case-Surveillance/COVID-19-Case-Surveillance-Public-Use-Data." (n.d.). Google Scholar. Retrieved December 7, 2020, from https://scholar.google.com/scholar?start=0&q=data.cdc.gov/Case-Surveillance/COVID-19-Case-Surveillance-Public-Use-Data&hl=en&as_sdt=0,11

[26] Ienca, M., & Vayena, E. (2020). On the responsible use of digital data to tackle the COVID-19 pandemic. Nature Medicine, 26(4), 463–464. https://doi.org/10.1038/s41591-020-0832-5

[27] Recommended Data Repositories | Scientific Data. (n.d.). Retrieved December 14, 2020, from https://www.nature.com/sdata/policies/repositories

[28] Search & Browse covid-19 | Page 1 of 4 | Data | Centers for Disease Control and Prevention. (n.d.). Retrieved November 19, 2020, from https://data.cdc.gov/browse?tags=covid-19